\documentclass[showpacs,twocolumn,showkeys,amsmath,amssymb,pra]{revtex4-1}

\usepackage{bm}
\usepackage{color}
\usepackage{graphicx}
\usepackage{placeins}
\newcommand{\rd}{{\mathrm d}}

\newcommand{\ri}{{\mathrm i}}

\newcommand{\ab}{a^{\phantom{\dagger}}}
\newcommand{\ad}{a^{\dagger}}
\begin{document}

\title[Entropy production]
      {Entropy production within a pulsed Bose-Einstein condensate}

\author{Christoph Heinisch}
\author{Martin Holthaus}

\affiliation{Institut f\"ur Physik, Carl von Ossietzky Universit\"at, 
	D-26111 Oldenburg, Germany}
                  
\date{April 27, 2016}

\begin{abstract}
We suggest to subject anharmonically trapped Bose-Einstein condensates to 
sinusoidal forcing with a smooth, slowly changing envelope, and to measure the 
coherence of the system after such pulses. In a series of measurements with 
successively increased maximum forcing strength one then expects an adiabatic 
return of the condensate to its initial state as long as the pulses remain 
sufficiently weak. In contrast, once the maximum driving amplitude exceeds a 
certain critical value there should be a drastic loss of coherence, reflecting 
significant heating induced by the pulse. This predicted experimental signature
is traced to the loss of an effective adiabatic invariant, and to the ensuing 
breakdown of adiabatic motion of the system's Floquet state when the many-body 
dynamics become chaotic. Our scenario is illustrated with the help of a 
two-site model of a forced bosonic Josephson junction, but should also hold 
for other, experimentally accessible configurations. 
\end{abstract} 




\maketitle 


\section{Introduction}
\label{sec:1}

In 1974 an influential series of experiments on the microwave-induced 
multiphoton ionization of highly excited Hydrogen atoms was initiated by
J.~E.~Bayfield and P.~M.~Koch~\cite{BayfieldKoch74}. Sending a beam of fast
Hydrogen atoms with principal quantum numbers ranging from $n = 63$ to 
$n = 69$ through an X-band microwave cavity and measuring the resulting
ionization probabilities as a function of the peak amplitude of the applied
oscillating electric field for a frequency of $9.9$~GHz, these authors 
observed the onset and saturation of ionization with increasing field strength,
although, formally, the energy of about 76 microwave photons was required 
to reach the continu\-um states from the initial state $n = 66$. Remarkably,
considerable ionization was observed even when the peak microwave amplitude
was small by comparison with the static electric field needed to ionize
the atom.

It was soon realized that the outcome of these early experiments is well
described by a classical approach, based on the evolution of a representative
set of classical trajectories in phase space~\cite{LeopoldPercival78}. Since
a classical anharmonic oscillator subjected to strong periodic driving 
naturally gives rise to chaotic dynamics, this inevitably led to the question 
how the chaotic behavior exhibited by the corresponding classical model is 
reflected in the ionization process of the real, quantum mechanical Hydrogen 
atom. Subsequent theoretical work on this question has led to remarkable 
insights. On the one hand, Casati {\em et al.\/} have predicted the existence 
of a certain critical micro\-wave field strength, dubbed quantum delocalization 
border, above which the quantum wave packet should delocalize, and strong 
excitation and ionization should take place~\cite{CasatiEtAl87}. On the other, 
Bl\"umel and Smilansky have drawn attention to a qualitative change of the 
system's Floquet states at the ionization border~\cite{BlumelSmilansky87}. 
Significant further interest in the experiments was spurred by the 
hypothesis, derived from an analysis of the quantum kicked 
rotator~\cite{FishmanEtAl82,GrempelEtAl84}, that the classical-quantum 
correspondence might be broken in the high-frequency regime: Whereas the
energy of chaotic classical trajectories grows diffusively, such diffusive
energy growth, and hence ionization, should be suppressed in the quantum 
system by means of a mechanism closely related to the Anderson localization
of particles moving on a one-dimensional disordered lattice. Indeed, 
signatures of this high-frequency stabilization have been reported in later 
works~\cite{GalvezEtAl88,JensenEtAl91}. Moreover, a detailed investigation
has been made concerning the influence of classical resonances, and their 
quantum mechanical counterparts, on the observed ionization 
signal~\cite{KochvanLeeuwen95}.                     
  
An additional twist to the interpretation of these micro\-wave ionization
experiments is provided by the fact that they are {\em not\/} performed 
with strictly time-periodic driving: When the fast atoms enter a microwave 
re\-sonator, they experience a fringe field which, in their rest frame, 
corresponds to a driving amplitude which increases slowly on the time scale 
set by one microwave cycle~\cite{GalvezEtAl88}. This means that the state 
experimented with within the cavity is actually prepared when entering it: 
The initial Rydberg state can either be shifted adiabatically into the 
connected Floquet state, or undergo multiphoton transitions at avoided 
crossings of its quasienergy~\cite{BreuerHolthaus89}. In the latter case the 
precise form of the fringe fields, which translates into the slowly varying 
envelope of the pulse actually ``seen'' by the atoms, can strongly influence 
the experimental results. Therefore, the question whether or not the actual 
wave function of the microwave-driven Hydrogen atom is able to adiabatically 
follow the instantaneous Floquet states under the action of a pulse provides 
a key for understanding the ionization data.
      
In the present paper we suggest to transfer these previous single-particle 
experiments, which have shaped much of our current understanding of periodically 
driven quantum systems, to the many-body level, utilizing Bose-Einstein 
condensates subjected to sinusoidal driving with a smooth, slowly changing 
envelope. Again, the key question then is whether or not the condensate wave
function responds to such pulses in an adiabatic manner. As we will argue, 
there may be far-reaching conceptual analogies between these older microwave 
experiments and the ones proposed here, with the classical-quantum 
correspondence being replaced by the correspondence between the mean-field 
description of a driven condensate and its full many-body dynamics. We proceed 
as follows: In Sec.~\ref{sec:2} we review a convenient $N$-particle model 
system, and discuss the results of selected numerical calculations which 
illustrate its interaction with forcing pulses. In Sec.~\ref{sec:3} we then 
compare the $N$-particle dynamics to the predictions of a mean-field approach, 
emphasizing that the latter remains trustworthy only in the regular regime, 
which allows for adiabatic following of the many-body wave function to the 
driving amplitude. In the final Sec.~\ref{sec:4} we formulate our conclusions.

\section{Model calculations}
\label{sec:2}

\begin{figure}[t]
\begin{center}
\includegraphics[width = 1.0\linewidth]{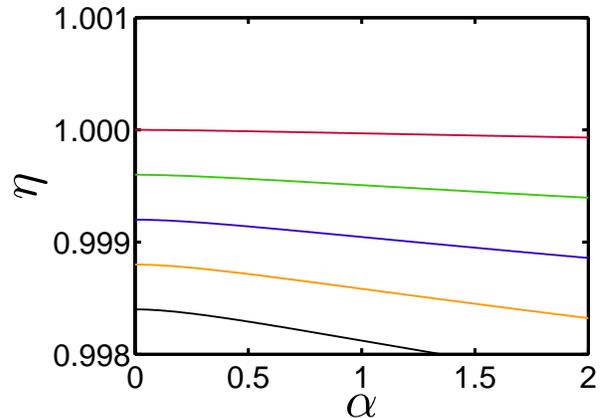}
\end{center}
\caption{Degree of coherence, as defined by Eq.~(\ref{eq:ETA}), for the 
	lowest five~eigenstates $n = 0,1,2,3,4$ (top to bottom) of the 
	model~(\ref{eq:UDJ}) with $N = 10\,000$ particles, vs.\ the scaled 
	interaction strength $\alpha = N\kappa/\Omega$. Observe that the 
	ground state $n = 0$ maintains $\eta_0 \approx 1$ to good approximation
	even up to $\alpha = 2.0$.} 
\label{F_1}	
\end{figure}

Our numerical studies are based on the familiar model of a bosonic Josephson 
junction~\cite{GatiOberthaler07} which supposes that the system consists of 
two sites connected by a tunneling contact, such that Bose particles sitting 
on the same site repel each other, while interaction between particles on 
different sites is neglected. Thus, the basic part of the model Hamiltonian 
reads~\cite{LMG65,ScottEilbeck86,MilburnEtAl97,Leggett01}
\begin{eqnarray}
	H_0 & = & -\frac{\hbar\Omega}{2} 
	\left( \ab_1\ad_2 + \ad_1\ab_2 \right) 
\nonumber \\ & & 
	+ \hbar\kappa \left( \ad_1\ad_1\ab_1\ab_1 
	+ \ad_2\ad_2\ab_2\ab_2 \right) \; .	
\label{eq:UDJ}
\end{eqnarray}
The bosonic operators $\ab_j$ and $\ad_j$, obeying the usual commutation 
relations
\begin{equation}
	\left[ \ab_j, \ab_k \right] = 0 	\; , \quad 
	\left[ \ad_j, \ad_k \right] = 0 	\; , \quad
	\left[ \ab_j, \ad_k \right] = \delta_{jk} \; , 
\end{equation}	
describe, respectively, the annihilation and creation of a particle at the
$j$th site ($j,k = 1,2$). The tunneling matrix element is written as 
$\hbar\Omega/2$, so that $\Omega$ denotes the single-particle tunneling 
frequency, while $2\hbar\kappa$ is the repulsion energy contributed by one 
pair of particles occupying a common site. Although the $N$-particle ground 
state of this system~(\ref{eq:UDJ}) is exactly equal to a coherent state, 
that is, to an $N$-fold occupied single particle state, for vanishing 
interparticle interaction only, it still remains a highly coherent 
condensate state even for finite values of the scaled interaction strength 
$\alpha = N\kappa/\Omega$~\cite{MazzarellaEtAl11,GertjerenkenHolthaus15b}. 
This is verified here by computing the one-particle reduced density matrix     
\begin{equation}
	\varrho_n = \left( \begin{array}{cc}
	\langle \ad_1 \ab_1 \rangle_n & \langle \ad_1 \ab_2 \rangle_n \\
	\langle \ad_2 \ab_1 \rangle_n & \langle \ad_2 \ab_2 \rangle_n 
		\end{array} \right) \; , 
\end{equation}		
where the expectation values $\langle \cdots \rangle_n$ are taken with respect 
to the $n$th energy eigenstate, from which one obtains the invariant 
\begin{equation}
	\eta_n  = 2 N^{-2} \, {\rm tr} \, \varrho_n^2 - 1 
\label{eq:ETA}
\end{equation}	
as a measure for the degree of coherence of that eigenstate: A pure condensate 
state gives ${\rm tr} \, \varrho_n^2 = N^2$, so that $\eta_n = 1$, whereas
a maximally fractionalized state yields ${\rm tr} \, \varrho_n^2 = N^2/2$,
amounting to $\eta_n = 0$~\cite{Leggett01}. In Fig.~\ref{F_1} we depict this
degree of coherence for the lowest five~eigenstates of a junction~(\ref{eq:UDJ})
occupied with $N = 10\,000$ Bose particles. Evidently, the ground state $n = 0$
remains highly coherent even up to $\alpha = 2.0$, and thus serves as a good 
condensate state.

Many authors have previously studied the dynamics of a bosonic Josephson 
junction~(\ref{eq:UDJ}) under the action of an external time-periodic force
with constant amplitude~\cite{HolthausStenholm01,SalmondEtAl02,SalasnichEtAl02,
MahmudEtAl05,WeissTeichmann08,GertjerenkenWeiss13}. In contrast, in order 
to monitor the system's adiabatic or non-adiabatic responses to forcing 
{\em pulses\/}, and thus to draw the parallels to the microwave ionization 
experiments reviewed in the Introduction, here we introduce a pulse-like, 
site-diagonal interaction of the form     
\begin{equation}
	H_1(t) = \hbar \mu(t) \sin(\omega t)
	\left( \ad_1\ab_1 - \ad_2\ab_2 \right) \; , 	
\end{equation}
where $\omega$ is the carrier frequency of the pulse, and $\hbar\mu(t)$ models 
its envelope, assumed to be slowly varying on the scale of $T = 2\pi/\omega$. 
In particular, we consider Gaussian envelopes 
\begin{equation}
	\mu(t) = \mu_{\rm max} \exp\left(-\frac{t^2}{2\sigma^2}\right)
\label{eq:PUL}
\end{equation}
with width parameter $\sigma$; their maximum strength $\mu_{\rm max}$ thus
carries the dimension of a frequency. The full model given by the total 
Hamiltonian   
\begin{equation}
	H(t) = H_0 +  H_1(t) 
\label{eq:TOT}
\end{equation}	
has also been employed in further recent investigations of pulsed many-Boson
dynamics~\cite{GertjerenkenHolthaus15b,GertjerenkenHolthaus15a,
HeinischHolthaus16}. Since the dimension of its Hilbert space figures merely
as $N+1$ when the driven junction hosts $N$ Bose particles, it allows one
to treat fairly large particle numbers with only moderate numerical effort.  

We now stipulate that initially, at time $t/T = -\infty$, the state of the 
system is given by the numerically determined condensate ground state of 
the undriven junction~(\ref{eq:UDJ}) for given, fixed particle number~$N$, 
and integrate the time-dependent Schr\"odinger equation with the 
Hamiltonian~(\ref{eq:TOT}) to determine the evolving $N$-particle state 
$| \psi(t) \rangle$. Here and in the following the reference time scale~$T$ 
is given by the carrier cycle time $T = 2\pi/\omega$; in practice, our 
numerical integrations cover the interval from $t_i = -10\,\sigma$ to 
$t_f = +10\,\sigma$.

\begin{figure}[t]
\begin{center}
\includegraphics[width = 1.0\linewidth]{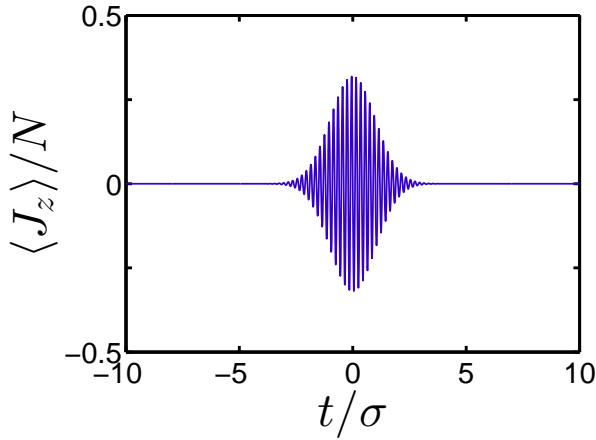}
\end{center}
\caption{Population imbalance~(\ref{eq:IMB}) of a driven system~(\ref{eq:TOT}) 
	consisting of $N = 100$ particles with scaled interaction strength 
	$N\kappa/\Omega = 0.95$, responding to a pulse~(\ref{eq:PUL}) 
	with carrier frequency $\omega/\Omega = 1.0$, maximum strength 
	$\mu_{\rm max}/\Omega = 0.60$, and width $\sigma/T = 5.0$, where 
	$T = 2\pi/\omega$ denotes the carrier cycle time. The initial 
	condensate state was the ground state of the undriven 
	junction~(\ref{eq:UDJ}). This is an example of an almost adiabatic 
	process, after which the system has returned almost completely to its 
	initial state.}   
\label{F_2}	
\end{figure}

Figure~\ref{F_2} shows the time-resolved population imbalance
\begin{equation}
	\langle J_z \rangle/N = \frac{1}{2N}
	\langle \psi(t) | \ad_1\ab_1 - \ad_2\ab_2 | \psi(t) \rangle
\label{eq:IMB}
\end{equation}
of a system comprising $N = 100$ particles with scaled interaction strength
$N\kappa/\Omega = 0.95$ while interacting with a pulse~(\ref{eq:PUL}) equipped 
with a carrier frequency $\omega/\Omega = 1.0$, moderate maximum strength
$\mu_{\rm max}/\Omega = 0.6$, and width $\sigma/T = 5.0$. Although such a 
pulse actually is comparatively short, so that the variation of the envelope 
during one single cycle~$T$ might not appear to be negligible, the system 
still is responding in an almost perfectly adiabatic manner: Adjusting itself 
to the ``slowly'' changing envelope, here the initial state is transformed 
almost perfectly into the connected instantaneous Floquet state belonging 
to the current driving amplitude, so that the system is able to return almost 
completely to its initial state with zero imbalance at the end of the 
pulse~\cite{BreuerHolthaus89,HeinischHolthaus16}.

\begin{figure}[t]
\begin{center}
\includegraphics[width = 1.0\linewidth]{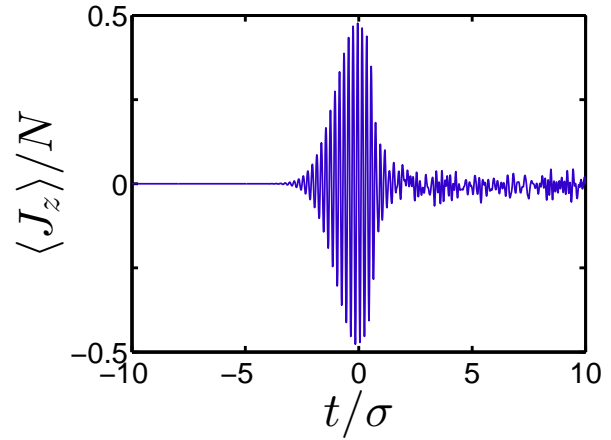}
\end{center}
\caption{As Fig.~\ref{F_2}, but with higher maximum driving amplitude
	$\mu_{\rm max}/\Omega = 0.90$. Here the system does not return to
	its initial state, indicating significant entropy production.}  
\label{F_3}	
\end{figure}

If the strength of the pulse is increased, the picture changes: In 
Fig.~\ref{F_3} we plot the imbalance for $\mu_{\rm max}/\Omega = 0.9$; 
all other parameters are the same as before. Now the system's response 
is no longer adiabatic: At the end of the pulse several eigenstates of the 
junction~(\ref{eq:UDJ}) are appreciably excited, leading to small, seemingly 
erratic fluctuations of the imbalance after the pulse is over.
  
In order to quantify this loss of adiabaticity with increasing pulse
strength in a more systematic manner, we determine the final occupation
probabilities 
\begin{equation}
	p_n = \big| \langle n | \psi(t_f) \rangle \big|^2 \; ,
\end{equation}
where $\{ | n \rangle \}$ denotes the eigenstates of the unperturbed
system~(\ref{eq:UDJ}), and compute the (dimensionless) von Neumann entropy 
generated by the pulse according to~\cite{PathriaBeale11}  	
\begin{equation}
	S = - \sum_n p_n \ln p_n \; .
\label{eq:ENT}
\end{equation}		 
Since there are $N+1$ different states, the maximum entropy 
$S_{\max} = \ln(N+1) \approx \ln N$ would result if all eigenstates of the 
junction were populated equally after the pulse, $p_n = 1/(N+1)$ for all 
$n = 0, \ldots, N$. The normalized entropy $S/\ln N$ therefore varies 
between zero and unity, with values close to unity indicating almost equal
distribution of the final state over the unperturbed energy eigenstates.

\begin{figure}[t]
\begin{center}
\includegraphics[width = 1.0\linewidth]{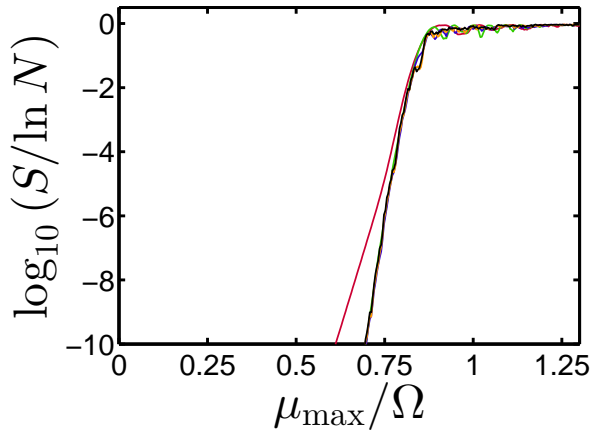}
\end{center}
\caption{Final entropy~(\ref{eq:ENT}) produced in a system with $N = 100$
	particles and scaled interaction strength $N\kappa/\Omega = 0.95$
	after pulses~(\ref{eq:PUL}) with frequency $\omega/\Omega = 1.0$
	and widths $\sigma/T = 5.0$ (red), $10.0$ (green), $20.0$ (blue), 
	$30.0$ (orange), and $40.0$ (black). Except for the shortest pulses
	with $\sigma/T = 5.0$, the curves almost fall on top of each other.} 
\label{F_4}	
\end{figure}

In Fig.~\ref{F_4} we depict the logarithm of this normalized final entropy
vs.\ the scaled maximum driving amplitude $\mu_{\rm max}/\Omega$, again for
$N = 100$, $N\kappa/\Omega = 0.95$, and $\omega/\Omega = 1.0$. Here we also 
consider several different pulse widths, $\sigma/T = 5$, $10$, $20$, $30$, 
and $40$, thereby allowing for a varying degree of adiabaticity. Except for 
the shortest pulses with $\sigma/T = 5$ all curves almost coalesce, giving
a quite consistent picture: Pulses with maximum scaled amplitude lower than 
$\mu_{\rm max}/\Omega \approx 0.65$ enable adiabatic following, and thus 
result in negligible entropy production. Then there is a transition regime,
extending from $\mu_{\rm max}/\Omega \approx 0.65$ to about      
$\mu_{\rm max}/\Omega \approx 0.85$, in which a slight increase of the 
maximum amplitude triggers a steep rise of the entropy. For still stronger
pulses one observes close-to-maximum entropy generation, corresponding to 
a significant heating of the initial condensate. 

\begin{figure}[t]
\begin{center}
\includegraphics[width = 1.0\linewidth]{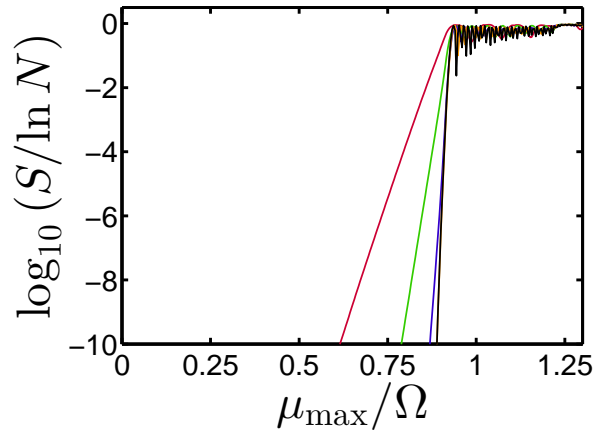}
\end{center}
\caption{As Fig.~\ref{F_3}, but with $N = 1000$ Bose particles.} 
\label{F_5}	
\end{figure}

\begin{figure}[t]
\begin{center}
\includegraphics[width = 1.0\linewidth]{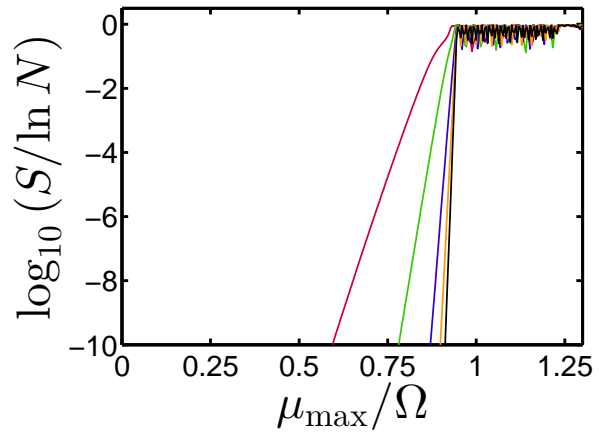}
\end{center}
\caption{As Fig.~\ref{F_3}, but with $N = 10\,000$ Bose particles.} 
\label{F_6}	
\end{figure}

In Figs.~\ref{F_5} and \ref{F_6} we repeat these investigations for $N = 1000$ 
and $N = 10\,000$ particles, respectively, keeping the scaled interaction 
strength $N\kappa/\Omega = 0.95$ fixed, so that an increase of the particle 
number~$N$ is accompanied by a reduction of the bare interparticle interaction 
strength~$\hbar\kappa$~\cite{GertjerenkenHolthaus15b}. Again one observes an 
adiabatic regime and a regime of close-to-maximum entropy production, separated
by a transition regime, but there is an additional feature: With these larger
particle numbers the system is able to discern the different pulse widths, 
meaning that the onset of entropy production shifts to higher maximum 
amplitudes when the pulse is made longer. On the other hand, the onset of 
the regime of maximum heating appears to be more or less independent of the 
pulses' length.

These numerical findings bear a striking similarity to the microwave 
ionization experiments~\cite{BayfieldKoch74} referred to in the Introduction: 
There one finds a ``delocalization border''~\cite{CasatiEtAl87} above which
a microwave pulse leads to significant ionization of a highly excited
Hydrogen atom; here one observes a ``heating border'' above which a driving
pulse leads to significant entropy generation in the model defined by the
Hamiltonian~(\ref{eq:TOT}). This model may be somewhat simple, and not capture
all aspects of the exact many-body dynamics of experimentally feasible driven 
bosonic Josephson junctions. Nonetheless, we surmise that its main constitutive 
element, the existence of a heating border, is a generic feature of pulsed 
Bose-Einstein condensates also in typical experimental configurations which 
do not lend themselves to exact numerical many-body calculations.

\section{The mean-field to $N$-particle correspondence}
\label{sec:3}

Within a mean field-type approach the dynamics of Bose-Einstein condensates 
are described by the time-dependent nonlinear Gross-Pitaevskii equation for 
the condensate's macroscopic wave function~\cite{Pitaevskii61,Gross63,
PethickSmith08,PitaevskiiStringari03}. In the case of the driven two-site
system governed by the Hamiltonian~(\ref{eq:TOT}) this equation is cast
into the form~\cite{GertjerenkenHolthaus15b}  
\begin{eqnarray}
	\ri \frac{\rd}{\rd \tau} c_1(\tau) & = & -\frac{1}{2} c_2(\tau)
	+ 2\alpha | c_1(\tau) |^2 c_1(\tau)
\nonumber \\	& &	
	+ \frac{\mu(\tau)}{\Omega}\sin\left(\frac{\omega}{\Omega}\tau\right)
	c_1(\tau) \; ,
\nonumber \\
	\ri \frac{\rd}{\rd \tau} c_2(\tau) & = & -\frac{1}{2} c_1(\tau)
	+ 2\alpha | c_2(\tau) |^2 c_2(\tau)
\nonumber \\	& &
	- \frac{\mu(\tau)}{\Omega}\sin\left(\frac{\omega}{\Omega}\tau\right)
	c_2(\tau) \; ,		
\label{eq:GPE}
\end{eqnarray}		
where $\tau = \Omega t$ is a dimensionless time variable, and
\begin{equation}
	\alpha = \frac{N\kappa}{\Omega}
\end{equation}
denotes the scaled interparticle interaction strength already employed 
in the previous section. The squared amplitudes $\big| c_j(\tau) \big|^2$ 
are to be interpreted as the expected fraction of particles occupying the 
$j$th~site at time~$\tau$ ($j = 1,2$). A detailed 
derivation~\cite{GertjerenkenHolthaus15b} of these coupled nonlinear 
equations~(\ref{eq:GPE}) which is not based on a coherent-states approach, 
but rather compares the evolution of an $N$-particle system to that of 
subsidiary systems obtained by the removal of one particle, reveals that 
they indeed do provide a faithful image of the exact $N$-particle dynamics 
generated by the Hamiltonian~(\ref{eq:TOT}) in the limit of large~$N$,
provided their solutions remain {\em regular\/}. In this case the dynamics in
Fock space are stiff, in the sense that the subsidiary $(N-1)$-particle states 
remain closely related to the actual $N$-particle state while evolving in
time. If, however, the solutions to Eqs.~(\ref{eq:GPE}) become {\em chaotic\/},
their direct link to the $N$-particle level is lost: In that latter case the 
exact $N$-particle state, when responding to the external drive, acquires a 
level of complexity which amounts to a destruction of the macroscopic wave 
function, so that its description in terms of a Gross-Pitaevskii approach 
becomes pointless~\cite{GertjerenkenHolthaus15b,GertjerenkenHolthaus15a}.

\begin{figure}[t]
\begin{center}
\includegraphics[width = 1.0\linewidth]{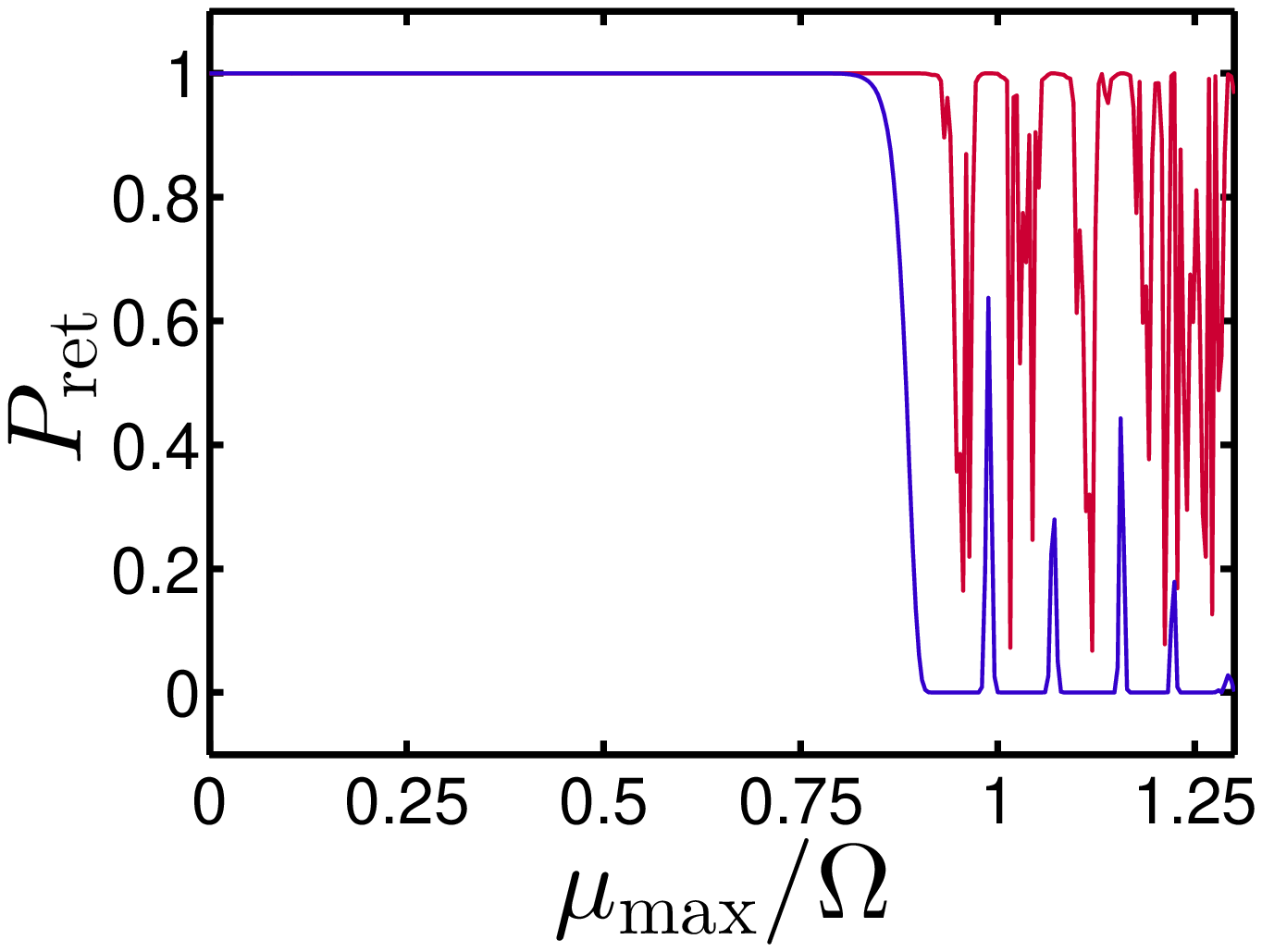}
\includegraphics[width = 1.0\linewidth]{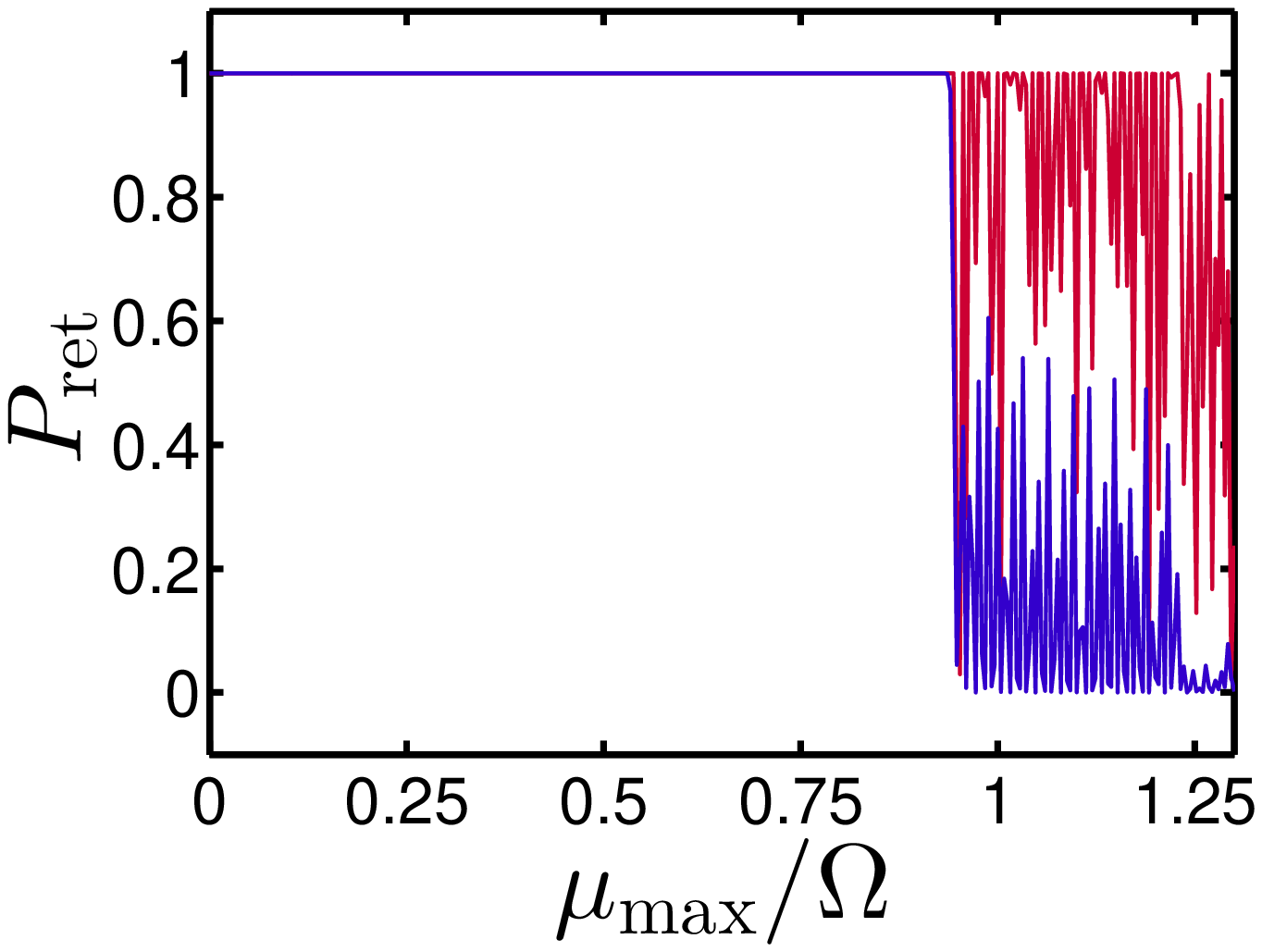}
\end{center}
\caption{Mean-field return probabilities~(\ref{eq:PMF}) (red) for systems
	with scaled interaction strength $\alpha = 0.95$ responding to pulses 
	with frequency $\omega/\Omega = 1.0$ and width $\sigma/T = 5.0$ 
	(upper panel) or $\sigma/T = 40.0$ (lower panel). These mean-field 
	data are compared to the corresponding quantum return 
	probabilities~(\ref{eq:PNP}) (blue), as computed for $N = 10\,000$ 
	particles.}   
\label{F_7}	
\end{figure}

We now consider the initial condition   	
\begin{equation}
	\left( \begin{array}{cc} 
		c_1(\tau_i) \\ c_2(\tau_i) 
	\end{array} \right) = \frac{1}{\sqrt{2}}
	\left( \begin{array}{cc}
		1 \\ 1
	\end{array} \right) \; ,	 	
\end{equation}
as corresponding to the condensate ground state $| 0 \rangle$ of the 
undriven junction~(\ref{eq:UDJ}), and utilize the Gross-Pitaevskii 
equation~(\ref{eq:GPE}) for computing the mean-field return probabilities
\begin{equation}
	P_{\rm ret}^{\rm mf} = \big|
	c_1^*(\tau_i) c_1(\tau_f) + c_2^*(\tau_i) c_2(\tau_f)
	\big|^2 
\label{eq:PMF}	
\end{equation}
for the same pulses as studied before. In Fig.~\ref{F_7} we display such
mean-field return probabilities for short and long pulses, $\sigma/T = 5$ 
(upper panel) and $\sigma/T = 40$ (lower panel), in comparison with the 
corresponding full $N$-particle quantum return probabilities
\begin{equation}
	P_{\rm ret}^{N} = \big| \langle 0 | \psi(\tau_f) \rangle \big|^2	
\label{eq:PNP}
\end{equation} 
obtained for $N = 10\,000$. For the shorter pulses the Gross-Pitaevskii 
treatment slightly overestimates the ``critical'' maximum amplitude
$\mu_{\rm max}/\Omega$ at which the sudden drop of the $N$-particle return
probability indicates the loss of adiabaticity, in line with a substantial 
entropy production. For the longer pulses, which should favor adiabatic
motion, the onset of a chaotic mean-field return pattern agrees very well 
with the quantum heating border. We conclude that the comparatively simple 
Gross-Pitaevskii equation, although it is lacking theoretical justification 
in the entropy-generation regime where no macroscopic wave function exists, 
is quite capable of determining the heating border itself.

\FloatBarrier
\section{Discussion}
\label{sec:4}

The field of ``quantum chaos'', which by now has led to a deepened 
understanding not only of the classical-quantum correspondence, but also 
of generic quantum dynamics itself, has previously concentrated on 
single-particle systems, chosen to be sufficiently simple to allow for 
full-fledged numerical analysis on both the classical and the quantum 
level~\cite{Gutzwiller91,Haake10,Stoeckmann07}. Externally driven Bose-Einstein
condensates now offer attractive prospects for extending these developments to 
an even more challenging many-body context. As long as such condensates respond 
to the drive in a ``regular'' manner, their macroscopic wave function remains 
preserved, possibly corresponding to a macroscopically occupied single-particle
Floquet state. If, however, one encounters the realm of ``chaotic'' dynamics, 
the many-body state no longer remains entropyless, but necessarily has to 
carry a lot of information, and therefore to acquire a high level of 
complexity which is not accessible to a mean-field ansatz. 

The mean-field dynamics associated with a periodically driven bosonic two-mode 
Josephson junction~(\ref{eq:UDJ}) are exactly equivalent to those of a driven 
nonlinear classical pendulum~\cite{GertjerenkenHolthaus15b}. Therefore, in 
this exceptional case one may approach chaotic many-body dynamics by mapping
out the corresponding classical phase space, as done in a recent experiment
reported by Tomkovi\v{c} {\em et al\/.}~\cite{TomkovicEtAl15}. However, in 
more general cases a reduction to two-mode dynamics will not be feasible, and 
more versatile tools of investigation will be required. Then the concept of 
adiabatic following of driven many-body states to the respective instantaneous 
Floquet states, as explored in the context of ``avoided-level-crossing 
spectroscopy'' for driven Bose-Hubbard-like systems~\cite{EckardtHolthaus08}, 
suggests itself: Adiabatic following of a Bose-Einstein condensate to a smooth 
forcing pulse amounts to the preservation of an entropyless condensate state 
which can exist in the regular regime only, whereas the entropy production 
associated with the loss of adiabaticity signals the onset of chaos.

Thus, it should be a worthwhile enterprise to set up an experiment for 
investigating the transition from regular, mean field-like condensate dynamics 
to chaotic many-body dynamics in systematic detail along the lines suggested 
in the present proposal: Start from a Bose-Einstein condensate in a trapping
configuration which allows for the application of well-defined forcing pulses. 
Then subject the condensate to a pulse with a slowly varying, smooth envelope, 
and measure the coherence of the final state after the pulse by recording, 
{\em e.g.\/}, the condensate fraction through time-of-flight absorption imaging. 
As long as the maximum pulse strength does not reach the chaotic regime, the 
system's response should be adiabatic. Thus, the condensate will return almost 
fully to its initial state after the pulse, with negligible production of 
entropy. If, however, the driving amplitude crosses the border to chaos, 
adiabatic following is disabled, and the system undergoes manifold excitations 
during the pulse, corresponding to a high entropy production. Such dynamical 
heating of the condensate leads to its destruction, detectable after the pulse 
through the loss of coherence. Therefore, in configurations possessing a sharp 
chaos border, which may be achievable with anharmonic trapping potentials, one 
should observe a sharp drop of the final coherence when the maximum driving 
amplitude is gradually increased.

From a theoretician's perspective, the difference between ``regular'' and 
``chaotic'' dynamics of periodically driven quantum systems is reflected 
in the properties of their quasienergy spectra when these are considered for 
all instantaneous driving amplitudes encountered during the pulses, with 
smooth coarse-grained quasienergy eigenvalues in the regular regime enabling 
effectively adiabatic following of the corresponding Floquet states, whereas 
multiple avoided quasienergy crossings in the chaotic regime give rise to 
multiphoton-like Landau-Zener transitions effectuating the observed entropy 
growth~\cite{Holthaus16}. While the numerical computation of such quasienergy 
spectra, and the verification of adiabatic following of many-body Floquet 
states, is still feasible for simplified models such as the driven Josephson 
junction considered here~\cite{HeinischHolthaus16}, this goal will remain 
unachievable for realistic mesoscopic condensates consisting of $10^6$ 
particles, say. Hence, the experiments suggested in this work might break new 
ground in an area where numerical guidance is feasible on the mean-field level 
only: The question whether the educated guesses derived from the present 
model study actually hold water for very large~$N$ has to be decided in the 
laboratory. 
     
Such condensate experiments would constitute a natural extension of 
previous microwave ionization experiments with highly excited Hydrogen 
atoms~\cite{BayfieldKoch74}, in which a classical chaos border manifests
itself through the onset of strong ionization. While that ionization border 
would find its match in the heating border discussed here, the question 
whether driven condensates may also give rise to an analog of the Anderson-like
suppression of classical phase-space diffusion reported for the quantum kicked 
rotator~\cite{FishmanEtAl82,GrempelEtAl84} appears to be open.

\begin{acknowledgments}
We are grateful for CPU time granted to us on the HPC cluster HERO, located 
at the University of Oldenburg and funded by the DFG through its Major 
Research Instrumentation Programme (INST 184/108-1 FUGG), and by the Ministry 
of Science and Culture (MWK) of the Lower Saxony State.
\end{acknowledgments}


\begin{thebibliography}{99}

\bibitem{BayfieldKoch74}
	J. E. Bayfield and P. M. Koch,
	Phys. Rev. Lett. {\bf 33}, 258 (1974).

\bibitem{LeopoldPercival78}
	J. G. Leopold and I. C. Percival,
	Phys. Rev. Lett.  {\bf 41}, 944 (1978).
	
\bibitem{CasatiEtAl87}
	G. Casati, B. V. Chirikov, D. L. Shepelyansky, and I. Guarneri, 
	Phys. Rep. {\bf 154}, 77 (1987).
	
\bibitem{BlumelSmilansky87}
	R. Bl\"umel and U. Smilansky, 
	Z. Phys. D {\bf 6}, 83 (1987). 

\bibitem{FishmanEtAl82}
	S. Fishman, D. R. Grempel, and R. E. Prange,
	Phys. Rev. Lett. {\bf 49}, 509 (1982)
		
\bibitem{GrempelEtAl84}		
	D. R. Grempel, R. E. Prange, and S. Fishman,	
	Phys. Rev. A {\bf 29}, 1639 (1984).	
		
\bibitem{GalvezEtAl88}
	E. J. Galvez, B. E. Sauer, L. Moorman, P. M. Koch, and D. Richards,
	Phys. Rev. Lett. {\bf 61}, 2011 (1988).	

\bibitem{JensenEtAl91}
	R. V. Jensen, S. M. Susskind, and M. M. Sanders,
	Phys. Rep. {\bf 201}, 1 (1991).
	
\bibitem{KochvanLeeuwen95}	
	P.M. Koch and K.A.H. van Leeuwen,
	Phys. Rep. {\bf 255}, 289 (1995).	

\bibitem{BreuerHolthaus89}
	H. P. Breuer and M. Holthaus,
	Z. Phys. D {\bf 11}, 1 (1989).

\bibitem{GatiOberthaler07} 
	R. Gati and M. K. Oberthaler,
	J. Phys. B: At. Mol. Opt. Phys. {\bf 40}, R61 (2007).

\bibitem{LMG65} 
	H. J. Lipkin, N. Meshkov, and A. J. Glick,
	Nucl. Phys. {\bf 62}, 188 (1965).

\bibitem{ScottEilbeck86} 
	A. C. Scott and J. C. Eilbeck,
	Phys. Lett. A {\bf 119}, 60 (1986).

\bibitem{MilburnEtAl97} 
	G. J. Milburn, J. Corney, E. M. Wright, and D. F. Walls,
	Phys. Rev. A {\bf 55}, 4318 (1997).	
		
\bibitem{Leggett01}
	A. J. Leggett,
	Rev. Mod. Phys. {\bf 73}, 307 (2001). 

\bibitem{MazzarellaEtAl11}
	G. Mazzarella, L. Salasnich, A. Parola, and F. Toigo,
	Phys. Rev. A {\bf 83}, 053607 (2011).	

\bibitem{GertjerenkenHolthaus15b} 
	B. Gertjerenken and M. Holthaus,
	Annals of Physics {\bf 362}, 482 (2015).
		
\bibitem{HolthausStenholm01}
	M. Holthaus and S. Stenholm,
	Eur. Phys. J. B {\bf 20}, 451 (2001).
	
\bibitem{SalmondEtAl02}
	G. L. Salmond, C. A. Holmes, and G. J. Milburn,
	Phys. Rev. A {\bf 65}, 033623 (2002).
	
\bibitem{SalasnichEtAl02}
	L. Salasnich, A. Parola, and L. Reatto,
	J. Phys. B: At. Mol. Opt. Phys. {\bf35}, 3205 (2002).	
	
\bibitem{MahmudEtAl05}
	K. W. Mahmud, H. Perry, and W. P. Reinhardt,
	Phys. Rev. A {\bf 71}, 023615 (2005).				
	
\bibitem{WeissTeichmann08}
	C. Weiss and N. Teichmann,
	Phys. Rev. Lett. {\bf 100}, 140408 (2008).
	
\bibitem{GertjerenkenWeiss13}
	B. Gertjerenken and C. Weiss,
	Phys. Rev. A {\bf 88}, 033608 (2013).			
		
\bibitem{GertjerenkenHolthaus15a}	 	 
	B. Gertjerenken and M. Holthaus,
	EPL {\bf 111}, 30006 (2015).
		
\bibitem{HeinischHolthaus16}
	C. Heinisch and M. Holthaus,	
	J. Mod. Opt. (2016); 
	DOI: 10.1080/09500340.2016.1167263.	
		
\bibitem{PathriaBeale11}
	See, e.g., R. K. Pathria and Paul D. Beale,
	{\em Statistical Mechanics\/}. Third Edition
	(Elsevier / Butterworth-Heinemann, Oxford, 2011).
	
\bibitem{Pitaevskii61} 
	L. P. Pitaevskii,
	Zh. Eksp. Teor. Fiz. {\bf 40}, 646
	[Sov. Phys. JETP {\bf 13}, 451] (1961).	

\bibitem{Gross63} 
	E. P. Gross,
	J. Math. Phys. {\bf 4}, 195 (1963).
		
\bibitem{PethickSmith08} 
	C. J. Pethick and H. Smith,
	{\em Bose-Einstein Condensation in Dilute Gases\/}. Second Edition 
	(Cambridge University Press, Cambridge, 2008).		
	
\bibitem{PitaevskiiStringari03} 
	L. Pitaevskii and S. Stringari,
	{\em Bose-Einstein Condensation\/}
	(Clarendon Press, Oxford, 2003).
		
\bibitem{Gutzwiller91}
	M. C. Gutzwiller,
	{\em Chaos in Classical and Quantum Mechanics\/}	
 	(Springer-Verlag, New York Berlin Heidelberg, 1991).
	
\bibitem{Haake10}
	F. Haake,
	{\em Quantum Signatures of Chaos\/}. Third Edition
	(Springer-Verlag, Berlin Heidelberg, 2010).			

\bibitem{Stoeckmann07}
	H.-J. St\"ockmann,
	{\em Quantum Chaos -- An Introduction\/}
	(Cambridge University Press, Cambridge, 1999). 
	
\bibitem{TomkovicEtAl15}
	J. Tomkovi\v{c}, W. Muessel, H. Strobel, S. L\"ock, P. Schlagheck,
	R. Ketzmerick, and M. K. Oberthaler, 
	arXiv:1509.01809.

\bibitem{EckardtHolthaus08}
	A. Eckardt and M. Holthaus,
	Phys. Rev. Lett. {\bf 101}, 245302 (2008).	
	
\bibitem{Holthaus16} For a tutorial review, see:
	M. Holthaus, 
	J. Phys. B: At. Mol. Opt. Phys. {\bf 49}, 013001 (2016).
	
\end{thebibliography}
\end{document}